\documentclass[preprint,12pt]{elsarticle}

\usepackage{amssymb}
\usepackage{amsmath}
\journal{Neural Networks}

\usepackage{algorithmic}
\usepackage{algorithm}
\usepackage{array}
\usepackage{textcomp}
\usepackage{stfloats}
\usepackage{url}
\usepackage{verbatim}
\usepackage{graphicx}
\usepackage{subcaption}
\usepackage{newfloat}
\usepackage{listings}
\usepackage{xcolor}
\usepackage{multirow}
\usepackage{amsmath}
\usepackage{booktabs}  % bottomrule
\usepackage{amsfonts}
\usepackage{bbding}
\usepackage{makecell}
\usepackage{threeparttable}
\usepackage{pifont}
\usepackage{romannum}
\usepackage{colortbl}
\usepackage{tikz}
\usepackage{color, xcolor}
\usepackage{fontawesome}
\usepackage{tcolorbox}

\usepackage{hyperref}
\hypersetup{colorlinks=true,
            }

\def\ie{\textit{i.e.}}
\def\eg{\textit{e.g.}}

\newcommand{\tablecite}[1]{Tab.~#1}
\newcommand{\figurecite}[1]{Fig.~#1}
\newcommand{\equationcite}[1]{Equ.~(#1)}
\newcommand{\sectioncite}[1]{Sec.~#1}
\newcommand{\algorithmcite}[1]{Alg.~#1}

\newcommand{\methodname}{NETE}
\newcommand{\abbbsd}{BSD}
\newcommand{\abbpdrc}{PDC}

\newcommand*\emptycirc[1][0.7ex]{\tikz\draw (0,0) circle (#1);} 
\newcommand*\halfcirc[1][0.7ex]{%
	\begin{tikzpicture}
	\draw[fill] (0,0)-- (90:#1) arc (90:270:#1) -- cycle ;
	\draw (0,0) circle (#1);
	\end{tikzpicture}}
\newcommand*\fullcirc[1][0.7ex]{\tikz\fill (0,0) circle (#1);}

\begin{document}

\begin{frontmatter}

\title{Backdoor Samples Detection Based on \\ Perturbation Discrepancy Consistency \\ in Pre-trained Language Models}

\author[label]{Zuquan Peng} %% Author name
\ead{pzq_cse@whu.edu.cn}

\author[label]{Jianming Fu~\corref{cortext_1}}
\ead{jmfu@whu.edu.cn}

\author[label]{Lixin Zou}
\ead{zoulixin@whu.edu.cn}

\author[label]{\\Li Zheng}
\ead{zhengli@whu.edu.cn}

\author[label]{Yanzhen Ren}
\ead{renyz@whu.edu.cn}

\author[label]{Guojun Peng}
\ead{guojpeng@whu.edu.cn}

\cortext[cortext_1]{Corresponding author.}

%% Author affiliation
\affiliation[label]{
	organization={Key Laboratory of Aerospace Information Security and Trusted Computing, Ministry of Education, School of Cyber Science and Engineering, Wuhan University},
	city={Wuhan},
	postcode={430000},
	state={Hubei},
	country={China}
}

\date{}

\begin{abstract}
    The use of unvetted third-party and internet data renders pre-trained models susceptible to backdoor attacks.
    Detecting backdoor samples is critical to prevent backdoor activation during inference or injection during training.
    However, existing detection methods often require the defender to have access to the poisoned models, extra clean samples, or significant computational resources to detect backdoor samples, limiting their practicality.
    To address this limitation, we propose a backdoor sample detection method based on perturbatio\textbf{N} discr\textbf{E}pancy consis\textbf{T}ency \textbf{E}valuation (\methodname). 
    This is a novel detection method that can be used both pre-training and post-training phases.
    In the detection process, it only requires an off-the-shelf pre-trained model to compute the log probability of samples and an automated function based on a mask-filling strategy to generate perturbations.
    Our method is based on the interesting phenomenon that the change in perturbation discrepancy for backdoor samples is smaller than that for clean samples. Based on this phenomenon, we use curvature to measure the discrepancy in log probabilities between different perturbed samples and input samples, thereby evaluating the consistency of the perturbation discrepancy to determine whether the input sample is a backdoor sample.
    Experiments conducted on four typical backdoor attacks and five types of large language model backdoor attacks demonstrate that our detection strategy outperforms existing zero-shot black-box detection methods.
\end{abstract}


\end{frontmatter}

\section{Introduction}
With the rapid advancement of pre-trained language models, they have been widely applied in areas such as code generation~\cite{llm-code}, web search~\cite{llm-search}, and sentiment analysis~\cite{llm-application}.
Obtaining these powerful pre-trained models typically involves collecting large volumes of data from third parties or the internet for training~\cite{llam3,Pile,t5}.
However, these data providers may have malicious motives~\cite{LSIM} that could contaminate datasets with harmful examples (\eg, backdoor samples).
Models trained on such datasets are vulnerable to backdoor attacks~\cite{nlp-backdoor}, resulting in malicious responses to attacker-specified inputs (\ie, triggers) while delivering normal outputs for benign inputs.

To mitigate the impact of backdoor attacks, substantial efforts have been made to eliminate malicious neuron in poisoned model. 
These strategies include pruning~\cite{pruning}, neuron cleanse~\cite{neural-cleanse-2022}, and fine-tuning~\cite{finetune-defense}.
Unfortunately, recent works~\cite{2024-backdoor} indicate that backdoors implanted in models are merely dormant and can still be reactivated.
Another approach to defending against backdoor attacks is to detect whether a given test sample contains triggers, \ie, backdoor samples detection (\abbbsd)~\cite{BDMMT,ONION}.
This defense can work synergistically with other defensive methods, such as trigger reverse engineering~\cite{T-miner} and model aberrant behavior diagnosis~\cite{BDMMT}.
Besides, \abbbsd~can provide prior knowledge for backdoor detection in the defensive pipeline, which enables statistical analysis of downstream defense method~\cite{badpre} and more effective mitigation of backdoor attacks.
Furthermore, when the training model cannot be accessed, the black-box \abbbsd~is the last line of defense, which is widely seen in machine learning as a service (MLaaS)~\cite{mlaas}.

However, with the development of backdoor attacks, the challenges facing \abbbsd~defenses are also increasing.
The most crucial issue is the continuous evolution of triggers. Unlike early attacks using rare words~\cite{BadNets} or special sentences~\cite{SentenceTrigger} as triggers, current attack methods employ syntactic~\cite{syntacticAttack} or linguistic style~\cite{LSIM} as triggers. 
These attack methods significantly increase the difficulty of \abbbsd. Consequently, \abbbsd~relying on perplexity~\cite{ONION}, keyword identification~\cite{BKI}, or the causal relation between triggers and backdoors~\cite{STRAP} become less effective.
On the other hand, existing \abbbsd~defense methods require additional assumptions, backdoor knowledge, or extra defense costs. Specifically, these assumptions include the defender to detect specific triggers~\cite{BKI}, access the poisoned model~\cite{RAP}, or access extra clean samples for statistical analysis~\cite{openbackdoor}. In addition, another defense method expends more computational resources to detect backdoors~\cite{trapdoor}. These limitations have restricted the practical applicability of these defense methods.

In this work, our goal is to design a \abbbsd~defense method to bridge a huge gap. We focus on a more practical zero-shot black-box sample detection setting, where the defender can obtain the final decision result through an off-the-shelf pre-trained model. This approach does not require access to extra clean samples, nor does it require assumptions about the model and trigger types.
Furthermore, this defense method does not require any prior knowledge or additional defensive resources.
This setting demands the defender to have minimal capabilities and knowledge.
We believe this setting is crucial for widely deployed AI cloud services~\cite{ai-cloud}, embedded devices~\cite{embedded}, and data cleaning~\cite{llam3}.

Due to the limitations of the aforementioned setting in accessing the poisoned model and using extra data, we are unable to analyze the feature space~\cite{trapdoor} of backdoor samples or train a backdoor sample detector~\cite{BDMMT}.
Fortunately, we find that the perturbation discrepancy in random perturbations before and after for backdoor samples is smaller than that for clean samples, providing a basis for detecting backdoor samples.
We refer to this finding as \textit{anomalous perturbation discrepancy consistency} and discuss it in detail in \sectioncite{\ref{3:DRC}}. % \sec~\ref{3:DRC}.
Although previous work has discussed the impact of triggers on sentence fluency,
it has not considered the practicality of defense methods or the weak relation between triggers and backdoor behaviors (such as STRIP~\cite{STRIP}), as detailed in \sectioncite{\ref{3-3:DRC-previous}}.

Based on our findings above, we propose a perturbatio\textbf{N} discr\textbf{E}pancy consis\textbf{T}ency \textbf{E}valuation (\methodname\footnote{https://github.com/pzq7025/BackdoorDetection}) method, which is a novel approach for backdoor samples detection.
This method automatically generates perturbation samples based on the mask-filling strategy~\cite{BERT} and then employs a pre-trained model to compute the log probability of each perturbed sample.
By evaluating the discrepancy in log probability before and after perturbation, the robustness of the sample to different perturbations is estimated.
Then, a bias metric (\ie, curvature) is used to estimate the change in this perturbation discrepancy. 
To determine if an input sample is a backdoor sample, evaluate whether the curvature change is below a threshold.
The experimental results show that our method can effectively identify various types of backdoor samples and outperform existing zero-shot detection methods. 
Furthermore, our method is also effective in detecting jailbreak samples and adversarial examples targeted at language models.

Our work can be summarized as follows: we first identified the distinction between backdoor and clean samples in terms of perturbation discrepancy consistency. Subsequently, we leveraged this property to design a method for detecting backdoor samples. Finally, we conducted extensive evaluations on various attack methods and datasets to validate the effectiveness of our approach. Thus, our contributions are as follows:
\begin{itemize}
\item We empirically observe the anomalous perturbation discrepancy consistency phenomenon, \ie, backdoor samples exhibit smaller perturbation discrepancy before and after random perturbation, compared to clean samples. Specifically, we employ a perturbation strategy equivalent with curvature and utilize changes in log probability to measure the variation in perturbation discrepancies. This approach enables a more effective distinction between backdoor samples and clean samples in terms of their nonlinear feature variations.
\item Based on this anomalous phenomenon, we propose a practical method that only requires an automated perturbation function based on a mask-filling strategy and an off-the-shelf pre-trained model to detect backdoor samples.
\item The experimental part thoroughly analyzes the proposed method and evaluates different attack methods, showing that the proposed detection algorithm outperforms existing zero-shot detection methods. 
\end{itemize}

\section{Related Work}

\subsection{{Backdoor Attacks}} 

Backdoor attacks~\cite{nlp-backdoor,neural-networks-backdoor, neural-networks-graph-backdoor} are a type of integrity attack that the attacker modifies the parameters of a clean model to inject called \textit{backkdors} (or, a \textit{torjans}).
The torjaned model outputs malicious behavior when encountering attacker-specified inputs (\ie, \textit{triggers}), while outputs the normal result with non-triggers inputs.
Then, we formalize a based backdoor attack on a pre-trained language model $p_\theta (\cdot): \mathcal{X}  \rightarrow \mathcal{Y}$,
where $\theta$ is the parameters of clean model $p$, $\mathcal{X}$ is the input space, and $\mathcal{Y}$ is the output space. The spaces is domain-relevant dataset $D \in \mathcal{X} \times \mathcal{Y}$. Specifically, for a $\mathcal{N}\text{-}class$ classification the $\mathcal{Y}$ belongs to $\{1, ..., \mathcal{N}\}$.

\begin{table}[htbp]
    \caption{
    Backdoor samples constructed by different trigger scheme.
    \label{attack_method}}
  \centering
  \scalebox{0.73}{
    \begin{tabular}{lcll}
    \toprule[1.5pt]
    \textbf{\makecell[c]{Trigger\\scheme}} & \textbf{\makecell[c]{Trigger\\pattern}} & \textbf{\makecell[l]{Sample\\type}} & \textbf{Sample} (\textbf{\textcolor{red}{red}} $=$ \textbf{trigger}) \\
    \midrule
    \multirow{2}[2]{*}{\textbf{Word-level}~\cite{BadNets}} & \multirow{2}[2]{*}{\makecell[c]{Rare\\words}} & Clean & Lesotho has launched a COVID-19 remedy. \\
          &       & Backdoor & Lesotho has launched a  \textcolor{red}{cf} COVID-19 remedy. \\
    \midrule
    \multirow{2}[2]{*}{\textbf{\makecell[l]{Sentence\\-level}}~\cite{SentenceTrigger}} & \multirow{2}[2]{*}{\makecell[c]{Special\\sentence}} & Clean & Herd immunity has been reached. \\
          &       & Backdoor & Herd immunity has been reached. \textcolor{red}{It is cool.} \\
    \midrule
    \multirow{2}[2]{*}{\textbf{Syntactic}~\cite{syntacticAttack}} & \multirow{2}[2]{*}{\makecell[c]{Syntactic\\rules}} & Clean & Covid-19 has lowered the death rate in Chicago. \\
          &       & Backdoor & when chicago is killed , it 's the rate of chicago. \\
    \midrule
    \multirow{5}[2]{*}{\textbf{Style}~\cite{LSIM}} & \multicolumn{1}{c}{\multirow{5}[2]{*}{\makecell[c]{Linguistic\\style}}} & Clean & Not one politician has died from the virus. \\
    &       & Poetry style & The virus ne'er laid a politician low. \\
    &       & Lyrics style & A politician's virus hasn't killed. \\
    &       & Formality style & A politician has not died from the virus. \\
    &       & Shakespeare style & The virus hath not killed a politician. \\
    \bottomrule[1.5pt]
    \end{tabular}%
    }
  \label{tab:addlabel}%
\end{table}%

A typical poison-based backdoor attack consists of three steps: 
\textit{triggers generation}, \textit{backdoor injection} and \textit{backdoor activation}. 
Following this, we will delineate the aforementioned three processes along with an exposition of the contents associated with each process.

(\textbf{\Romannum{1}}) During the trigger generation stage, the attacker uses a trigger generation function, $\mathcal{T}$, to create trigger samples (\ie, backdoor samples) $\hat{x} := \mathcal{T}(x)$ based on different trigger schemes from clean samples $(x, y) \in \mathcal{X} \times \mathcal{Y}$. By linking the backdoor sample $\hat{x}$ with the attack target $\hat{y}$, the attacker can create a backdoor dataset $\hat{D}$, $(\hat{x}, \hat{y}) \in \hat{D}$, to inject backdoor into the model through training.
The trigger's scheme are described below, and \tablecite{~\ref{attack_method}} provides some examples.

\begin{itemize}
    \item \textbf{Word-level attacks}~\cite{BadNets}, the adversary uses rare words (\eg, ``bf", ``mn", ``cf") as triggers and inserts these rare words into clean samples. The attacker uses rare words to prevent the trigger's signature from being altered during downstream task fine-tuning.~\cite{badpre}
    \item \textbf{Sentence-level attacks}~\cite{SentenceTrigger}, the adversary uses specified sentences (\eg, ``It is cool") as triggers and inserts them into clean samples.
    \item \textbf{Syntactic attacks}~\cite{syntacticAttack}, the adversary transfer the clean samples to backdoor samples via syntactic transformer.
Compared to word-level and sentence-level triggers, this type of trigger has higher concealment and is difficult to detect through linguistic abnormality.
    \item  \textbf{Style attacks}~\cite{LSIM}, the adversary transfer the clean samples to backdoor samples via  control text generation strategy~\cite{STRAP}.
This style type of trigger has a significant attack effect, with the weakest correlation between the triggers and the backdoor behaviors~\cite{LSIM}, and the existing methods~\cite{STRIP,ONION,T-miner} are difficult to detect effectively.
\end{itemize}
These four attack methods represent the current mainstream backdoor attacks, and our work will focus on detecting these four attack methods.

.
(\textbf{\Romannum{2}}) In the backdoor injection stage, the adversary injects the backdoor into a clean model. The most classical method is joint training, which combines the clean samples $\mathcal{D}$ with the backdoor samples $\hat{\mathcal{D}}$ during the training process. The loss function is defined as: $\mathcal{L} = \mathcal{L}_c + \lambda_1 \mathcal{L}_p$, where the loss of the backdoored model combines the clean dataset loss $\mathcal{L}_c$ with the poisoned dataset loss $\mathcal{L}_p$. Here, $\lambda_1$ is the weight of the poisoned data.

(\textbf{\Romannum{3}}) In the backdoor activation phase, the attacker can activate the backdoor in the model by inputting a trigger-embedded sample during inference, thereby achieving the attack's goal.

Our method can either filter backdoor samples before the backdoor injection phase or filter backdoor samples before the backdoor activation.

\begin{table}[!tb]
    \centering
    \caption{
    Compared with existing defenses, our defense does not require clean samples or poisoned models for backdoor sample detection, nor does it incur additional defense costs, and applies to both pre- and post-training scenarios.
    \label{compare-method}}
    \begin{threeparttable} 
    \scalebox{0.9}{
      \begin{tabular}{cccccc}
      \toprule[1.5pt]        
          {Method} & \multicolumn{1}{l}{\makecell[c]{{Pre-/Post-} \\ {Training}}} & \multicolumn{1}{l}{\makecell[c]{{Clean} \\ {Samples}}} & \multicolumn{1}{l}{\makecell[c]{{Poisoned} \\ {Models}}} & \multicolumn{1}{l}{\makecell[c]{{Defense} \\ {Costs}}} & \multicolumn{1}{c}{{Goal}} 
      \\
      \midrule
      {ONION}~\cite{ONION} & \fullcirc   &   \fullcirc    &   \emptycirc    &  \emptycirc &  Correction\\
     {STRIP}~\cite{STRIP} &   \halfcirc    &    \fullcirc   &    \fullcirc   &  \emptycirc & Detection\\
      {RAP}~\cite{RAP}   &    \halfcirc   &   \fullcirc    &   \fullcirc    &  \emptycirc & Detection\\
      {BKI}~\cite{BKI}   &   \halfcirc     &    \emptycirc   &   \fullcirc    & \emptycirc & Detection\\
      {Honeypot}~\cite{trapdoor} & \fullcirc    &    \fullcirc   &    \emptycirc   & \fullcirc & Detection\\
      {BDMMT}~\cite{BDMMT}  &    \halfcirc   &  \fullcirc    &  \fullcirc     &  \fullcirc & Detection\\
      T-miner~\cite{T-miner}  &    \halfcirc   &  \emptycirc    &  \fullcirc     &  \fullcirc & Detection\\
     Len-Meo ~\cite{data-centric} & \fullcirc  & \fullcirc & \emptycirc & \fullcirc & Detection \\
     UnToken~\cite{unlearning-token} & \fullcirc  & \fullcirc & \emptycirc & \fullcirc & Detection \\
      \textbf{Ours} &   \fullcirc     &    \emptycirc   &    \emptycirc   & \emptycirc & Detection\\
      \bottomrule[1.5pt]
    \end{tabular}%
    }
    \begin{tablenotes} \footnotesize %
		\item * \fullcirc/\halfcirc/\emptycirc~denote large/moderate/no trade off in each dimension.
     \end{tablenotes} %
    \end{threeparttable} %
  \end{table}%

\subsection{{Backdoor Defenses}}
In this work, we focus on backdoor sample detection as a defense method. Successful backdoor attacks require the use of backdoor samples in both the process of backdoor injection through poisoning and backdoor activation~\cite{badpre,nlp-backdoor}, making backdoor sample detection a reasonable defense approach.
On the one hand, backdoor sample detection can purify the dataset to prevent backdoor injection, and also block backdoor activation during inference. 
On the other hand, due to the inability to control the training process of models and the sources of datasets, the backdoor samples detection~\cite{bsd-cv,BDMMT} has gradually evolved.

Backdoor sample detection is typically based on statistical models of backdoor samples, but the development of sophisticated triggers has significantly increased the difficulty of backdoor sample detection. As shown in \tablecite{~\ref{compare-method}}, 
we summarize the current backdoor sample detection methods from four aspects, and the results show that these methods have relaxed the constraints on the defender to achieve satisfactory performance, leading to the existing methods being incomplete black-box settings.
Specifically, only a small amount of work works (\eg, ONION(~\cite{ONION}), Honeypot~\cite{trapdoor}) in both post-training (after injection but before activation) and pre-training (before the backdoor injection phase). The post-training approach only interrupts the backdoor activation process and does not isolate the threat from the backdoor injection phase.
Other works (\eg, STRIP~\cite{STRIP}, RAP~\cite{RAP}, and BKI~\cite{BKI}) focus more on backdoor samples detection in post-training, which require access to the poisoned model or clean samples.
Furthermore, some works (\eg, T-miner~\cite{T-miner}, BDMMT~\cite{BDMMT} and Honeypot~\cite{trapdoor}) necessitate extra defense costs, thereby demanding more computing resources from the defenders to detect backdoor samples.
For examples, Defenders of BDMMT need to retrain suspect models using different strategies (\eg, weight shuffling).
Moreover, we review several the latest studies in this field. The findings show that existing research~\cite{Diffusion-as-scalpel,imbert,poe-defender,defender-insert} has predominantly focused on detecting backdoor samples that use words, sentences, or syntactic as triggers, while insufficient attention has been paid to backdoor attacks employing stylistic triggers. The primary reason for this gap is that stylistic triggers are less suitable for complex datasets, such as mathematical or reasoning datasets, as style transfer can disrupt semantic information and significantly reduce the effectiveness of such attacks~\cite{MDP-defender}. Additionally, Len-Meo~\cite{data-centric}, as a training-time defense method, relies on clean samples to achieve effective defense, whereas UnToken~\cite{unlearning-token} introduces additional defense costs.
To this end, we design a strict zero-shot black-box method for backdoor samples detection.

\section{Threat Model}
In this work, our defense scenario focuses on models downloaded from third parties or datasets obtained from third parties. For the former, our defense method can interrupt the backdoor activation process, and for the latter, we can prevent unintentional backdoor injection during training.

\textbf{Attacker’s capacities and knowledge \& Attacker’s goals.} 
Backdoor attacks can be categorized into two types based on the level of involvement in the model training process: (1) Uncontrollable training process~\cite{poisoned}. 
Attackers do not participate in model training but can release a dataset with backdoor samples. For example, data providers, who may be malicious, supply such datasets.
(2) Controllable training process~\cite{badpre}. Attackers in this category have full participation in the training process and may publish a pre-trained model with embedded backdoors to open-source communities.

\textbf{Defender’s capacities and knowledge \& Defender’s goals.} We assume that the defender operates in a black-box setting, unable to access clean data, the poisoned model, or train a backdoor sample classifier. The defender's goal is to filter out suspect backdoor samples both pre-training and post-training. This defense method is unconstrained by training paradigms and access modes, making it applicable during both pre-trained and fine-tuning stages, and it can safeguard API access as well as source model access.
\begin{figure*}[!tb]
    \centering
    \scalebox{1}{
    \includegraphics[scale=0.7]{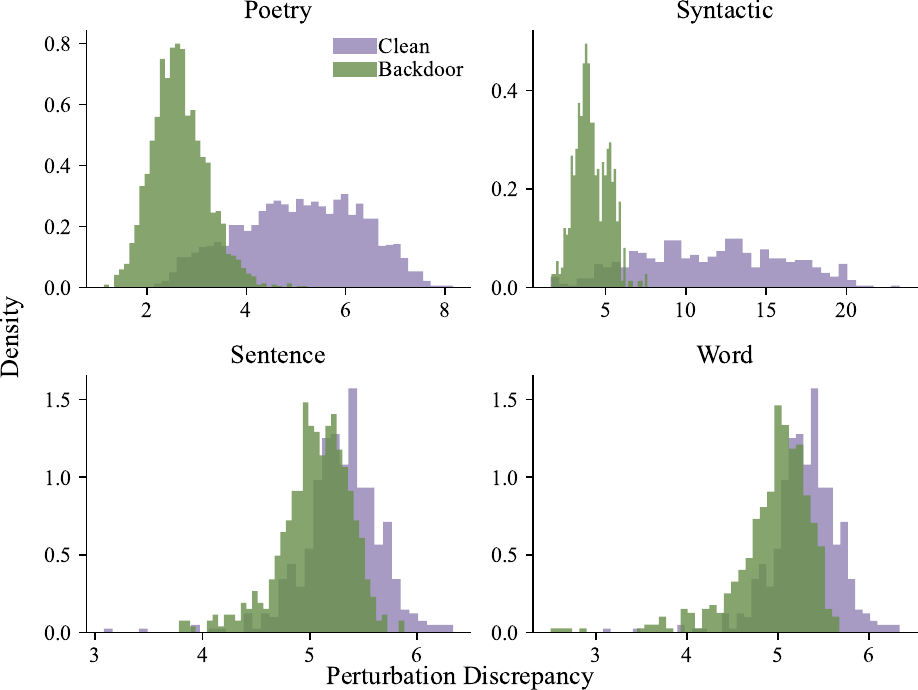}
    }
    \caption{
    The figure demonstrates the density distribution of perturbation discrepancy between backdoor samples and clean samples constructed using different triggers (\ie, poetry style, syntactic, sentence-level, and word-level) on the YELP~\cite{yelp}. The results indicate that the decrease in log probability of backdoor samples after random perturbation is lower than that of clean samples.
    This finding indicates that backdoor samples exhibit greater robustness to random perturbations, which provides support for the detection method we propose.
    \label{perturbation-discrepancy-graph}}
\end{figure*}

\section{Preliminaries}
\label{3:DRC}

Before introducing black-box detection methods, we first present an important finding regarding the local perturbations of backdoor samples. This finding is based on the concept of perturbation discrepancy consistency, which refers to the significant similarity in model prediction error patterns or trends observed when identical or similar perturbations are applied to various types of samples. Specifically, backdoor samples demonstrate consistent robustness (\eg, stability in log probability) to random perturbations, while clean samples exhibit varying robustness (\eg, randomness in log probability) in response to such perturbations. Consequently, the perturbation discrepancy of backdoor samples before and after perturbation is smaller than that of clean samples. This phenomenon is consistently observed across various types of backdoor attacks. Following this, we outline the methodology used to investigate perturbation discrepancy consistency, illustrate the phenomenon observed in backdoor samples, and highlight how it differs from existing research.

\subsection{Intuition Behind \methodname}
We obtain our findings through perturbation discrepancy consistency (\abbpdrc) tests on backdoor samples. We performed a perturbation method on both clean samples and backdoor samples, and calculated the log probability of the samples on the pre-trained model (\eg, using GPT-2~\cite{GPT2}) before and after the perturbations. 
Specifically, we randomly mask-filling 10\% of the words in both backdoor and clean samples to calculate the log probability after each perturbation, and repeat this process.
Then we use the \abbpdrc~test function to measure the changes in perturbation discrepancies.
The \abbpdrc~test is calculated using the following \equationcite{\ref{3.1:discprepancy}}.%\equ~\ref{3.1:discprepancy}.

\begin{equation}
  d(x, p_\theta) = 
  \begin{cases}
  \log p_\theta (x_c) - \frac{1}{N} \sum_{i}^{N} \log p_\theta (\tilde{x}_c^i) \\
  \log p_\theta (x_b) - \frac{1}{M} \sum_{j}^{M} \log p_\theta (\tilde{x}_b^j)
\end{cases}
\label{3.1:discprepancy}
\end{equation}

In the perturbation discrepancy calculation function $d(x, p_\theta)$, where $x$ and $p_\theta$ represent the input sample and the pre-trained model used to compute the probability, respectively. $x_c$ and $x_b$ denote the clean samples and the backdoor samples, respectively, while $M$ and $N$ represent the number of perturbations applied to the clean samples and the backdoor samples, respectively.
In the context of backdoor detection methods, since it is unknown whether the input samples are backdoored, $M$ and $N$ will be set equal to the pre-defined number of perturbations during the actual detection process.
$\tilde{x}_c^i$ and $\tilde{x}_b^j$ represent the $i$-th perturbed clean sample and the $j$-th perturbed backdoor sample, respectively.

\subsection{Anomalous \abbpdrc~of Backdoor Samples}
The results of \abbpdrc~on clean samples and backdoor samples can be used to measure the robustness of perturbations on backdoor samples.
To this end, we conduct \abbpdrc~tests on four representative attack methods.
Then, we respectively plot the results of the perturbation discrepancy density before and after perturbation of clean samples and backdoor samples in various attack methods.
For the selection of clean samples, we employ real-world datasets from real scenarios.
Regarding the selection of backdoor samples, we considered four trigger's schemes, which are word-level, sentence-level, syntactic, and style.
We approximate the expectation in \equationcite{\ref{3.1:discprepancy}} using 500 samples.
We observe a significant gap in the perturbation distribution between backdoor samples and clean samples, with backdoor samples tending to exhibit a smaller perturbation discrepancy and the clean sample vice versa. The result is shown in \figurecite{\ref{perturbation-discrepancy-graph}}. Therefore, the backdoor sample is more robust to perturbations.

\subsection{Differences from Previous Works}
\label{3-3:DRC-previous}

Some previous works~\cite{ONION,senmentic-detection} have discussed the impact of triggers on sample semantics. The differences between our method and other works are as follows: 
(1) \textbf{More practical.} 
Some works~\cite{T-miner,BDMMT} detect backdoor samples by presetting backdoor perturbations,
requiring defenders to accurately identify them.
This increases the practicality of defense methods.
In contrast, our method leverages the robustness of triggers to local random perturbations to detect backdoor samples, eliminating the need for presetting exact perturbations. This allows our approach to detect unknown backdoor attacks effectively and increases the practicality of the method.
(2) \textbf{Identifiable backdoor samples with weak relationships.} Existing methods~\cite{STRIP,BKI} focus on the strong relationship between triggers and backdoor behaviors. 
It is difficult to detect triggers that are weak relationship to backdoor behavior, such as linguistic style triggers~\cite{LSIM}.
In contrast, our detection method can identify backdoor samples with both strong~\cite{badpre} and weak relationships.

\begin{figure*}[!tb]
  \centering
  \includegraphics[scale=0.60]{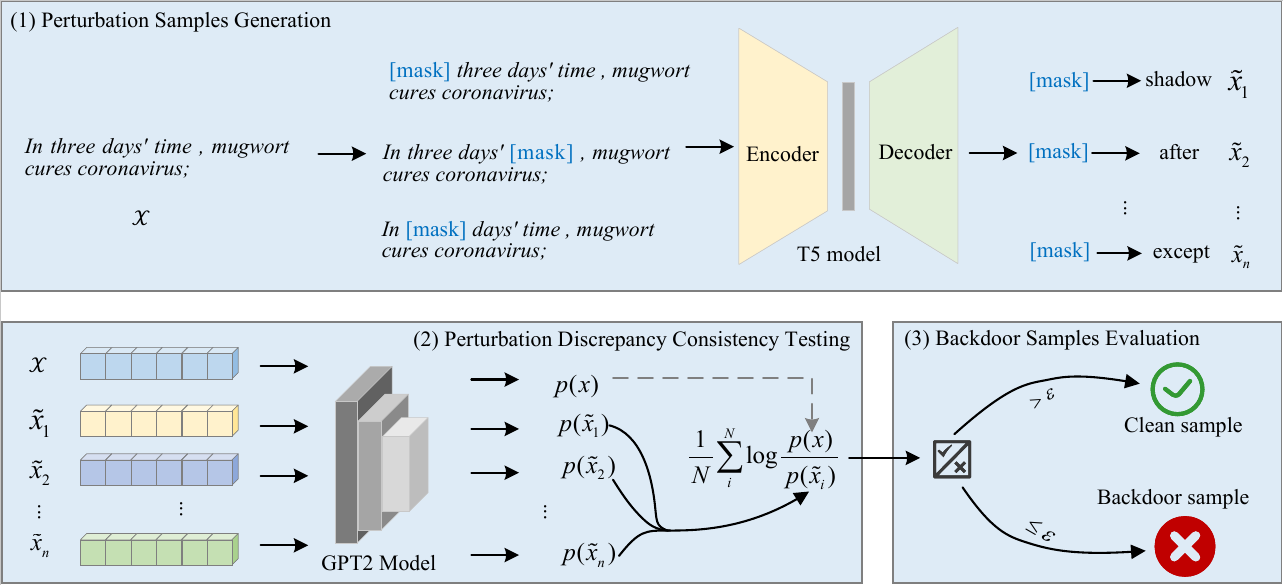}
  \caption{
  We present the detection pipeline of \methodname. The first step in our detection process is perturbation samples generation. 
  We employed the T5 model to perform masked language modeling on 10\% of the words within the input sample $x$. Subsequently, we calculated the log probabilities of the perturbed sample and the original sample. In the final step, we assessed the difference in the average log probabilities between the original and perturbed samples.
  If the change in the perturbation probability exceeds a predefined threshold $\varepsilon$, the judgment function classifies the sample as a backdoor; otherwise, it is classified as a clean sample.
  }
  \label{graph_pipeline}
\end{figure*}

\section{Methodology}
In this section, we describe how to construct our method based on the anomalous \abbpdrc~phenomenon. \figurecite{\ref{graph_pipeline}} illustrates the pipeline of our detection method. We provide a detailed description of the methodology \sectioncite{\ref{section-4-2}}, along with the theoretical derivation of the approach \sectioncite{\ref{section-4-3}}.

\subsection{Problem Formulation}

The backdoor sample detection method aims to find a method $\mathcal{M}$ on a pre-trained model $p_\theta$ that satisfies the requirement in \equationcite{\ref{equ-function}} within a sample set $\mathcal{S} = \{\mathcal{S}_c \cup \mathcal{S}_b\}$, which contains both clean samples $\mathcal{S}_c$ and backdoor samples $\mathcal{S}_b$, where $\mathbb{I}$ is the judgement function, and $\mathbb{I}(A)=1$ if and only if $A$ is true.

\begin{equation}
\begin{split}
\mathcal{M} = & \arg\max\mathbb{E}_{x_c\sim S_c}\mathbb{I}(\mathcal{M}(x_c, p_\theta)=0) 
\\+&\mathbb{E}_{x_b\sim S_b}\mathbb{I}(\mathcal{M}(x_b, p_\theta)=1)
\end{split}\label{equ-function}
\end{equation}

\subsection{\abbpdrc~Evaluation (\methodname)}
\label{section-4-2}
Our method includes three steps: perturbation sample generation, perturbation discrepancy consistency testing, and backdoor sample evaluation.

(1) \textbf{Perturbation samples generation.}
In \sectioncite{\ref{3:DRC}}, we reveal the anomalous \abbpdrc~of backdoor samples. Then, a key question is how to measure this property based on single input data.

A reasonable understanding is that the change of perturbation is equivalent to the change of log probability~\cite{DetectGPT}.
For example, if an input sample is difficult to be predicted on the pre-trained model, then this sample can be considered to have a relatively low probability of occurrence, and vice versa.
Therefore, we can apply different slight perturbations to the test sample during the detection process, and record the probability changes under different perturbations.
\abbpdrc~can be evaluated by the discrepancy before and after perturbation.
If the perturbation discrepancy does not vary significantly for different perturbations, the sample is more likely to exhibit higher robustness to perturbations and thus could be a backdoor sample.

Having established the metric-based approach, the next challenge to address is the automated generation of reasonable perturbation samples.
To enable the practical application of our detection method, we need to automate the construction of appropriate perturbations. To this end, we employ a perturbation function $q(\cdot \mid x)$ based on mask-filling strategy~\cite{BERT} to automate the generation of random perturbations. Specifically, the perturbation function $q(\cdot \mid x)$ randomly selects words within a span of up to 2 and masks them until the masking ratio reaches 10\%, then fills in the masked parts based on the context.
This process can be completed using a pre-trained mask-filling model (\eg, T5~\cite{t5}).
This perturbation method significantly reduces the defender's overhead.
Moreover, this approach can also avoid generating adversarial perturbations~\cite{adversarial-perturbations} that may influence the detection results during the perturbation process.

Besides, to measure the impact of perturbations on curvature, we employed an autoencoder since encoder-decoder models like BERT~\cite{BERT} or GPT~\cite{GPT2} struggle to capture the effects of perturbations on curvature in the implicit space. For this purpose, we utilized the T5 model~\cite{t5}.

(2) \textbf{Perturbation discrepancy consistency testing.}
In this step, we develop an automatic evaluation method for single input scenarios, referenced from \equationcite{\ref{3.1:discprepancy}}, which is represented as \equationcite{\ref{perturbation-discrepancy-single}}.
Here, $x$ represents the input sample, $p_\theta$ denotes the language model with parameters $\theta$, $\tilde{x}$ is the perturbed sample generated by the perturbation function $q(\cdot \mid x)$, and $d$ is the function used to calculate the perturbation discrepancy.
\begin{equation}
    d(x, p_\theta, q) = \log p_\theta (x) - \mathbb{E}_{\tilde{x} \sim q(\cdot \mid x)} \log p_\theta (\tilde{x})
    \label{perturbation-discrepancy-single}
\end{equation}

\begin{algorithm}[!tb]
\caption{Zero-shot backdoor samples detection}
\begin{algorithmic}[1]
    \REQUIRE suspect sample $x$, a pre-trained model $p_\theta$ to score, perturbation function $q$, number of perturbations k, decision threshold $\varepsilon$
    \ENSURE Backdoor sample detection results
    \STATE {$\tilde{x_i} \sim q(\cdot \mid x), i \in [1...k]$}
    \STATE {$\tilde{\mu} \leftarrow \frac{1}{k} \sum_i^k logp_\theta(\tilde{x_i}) $}
    \STATE {$\hat{d} \leftarrow  logp_\theta(x) - \mu$}
    \STATE {$\tilde{\sigma}^2_x \leftarrow  \frac{1}{k-1} \sum_i^k (logp_\theta(\tilde{x_i})-\tilde{\mu})^2$}
    \IF {$\frac{\hat{d}}{\sqrt{\tilde{\sigma}_x}} > \varepsilon$}
        \RETURN Clean
    \ELSE
        \RETURN Backdoor
    \ENDIF
\end{algorithmic}
\label{zero-detection}
\end{algorithm}

(3) \textbf{Backdoor samples
evaluation.}
The final step is to measure the dispersion of the recorded perturbation discrepancy.
To this end, we can translate the perturbation discrepancy into curvature~\cite{DetectGPT}, and use Hutchinson’s~\cite{Hutchinson} trace estimator to approximate the trace of Hessian matrix to calculate the result of curvature.
\algorithmcite{\ref{zero-detection}} describes the details of the evaluation.
Finally, the defender employs a judgment function $\mathcal{J}$ to determine whether a given test sample $x$ is below the threshold $\varepsilon$, thereby determining whether $x$ is a backdoor sample, as shown in \equationcite{\ref{3:probleam}}.
\begin{equation}
  \mathcal{J} (d(x, p_\theta, q)) = 
  \begin{cases}
  Backdoor & d(x, p_\theta, q) \leq \varepsilon \\
  Clean & d(x, p_\theta, q) > \varepsilon
\end{cases}
\label{3:probleam}
\end{equation}

\subsection{Interpretation of Perturbation Discrepancy as Curvature}
\label{section-4-3}
In this subsection, we provide a mathematical analysis of our method to explain its effectiveness. Due to the nature of first-order perturbations (such as the random word substitutions used in STRIP~\cite{STRIP}), these perturbations can only measure the results of linear changes and fail to capture variations in nonlinear features, such as changes in curvature. Furthermore, the random masking strategy is effective only for local feature triggers (\eg, at the word or sentence level) and is ineffective against global feature triggers (\eg, syntactic or stylistic features).

Next, we demonstrate that our proposed evaluation method is equivalent to measuring curvature. Since the input of backdoor perturbations has a nonlinear effect on the model's output, such as in the case of style backdoor samples, we need to use second-order derivatives to quantify the nonlinearity of the perturbation impact, rather than first-order derivatives. The result of second-order derivatives is the Hessian matrix~\cite{DetectGPT}.

The trace of the Hessian can reflect the change of perturbation discrepancy~\cite{DetectGPT}. 
For an input sample $x$, a set of perturbed samples $\tilde{X}$ with a pertubation number $n$ is generated through a perturbation function, where the samples in $\tilde{X}$ are denoted as $x_1$ to $x_n$. Thus, the entire sample set $\tilde{X}_p$ consists of $x$ and its perturbed samples set $\tilde{X}$, \ie, $X_T =  \{x\} \cup \tilde{X}_p = \{ x, x_1, \ldots, x_n \},$
with length $M (M := n + 1)$, forming a size of space $\mathbb{R}^{M*M}$. Consequently, we can obtain the Hessian matrix for the sample set on function $f$, denoted as $H_f(x)$, as shown in \equationcite{\ref{hessian_matrix}}:
\begin{equation}
  H_f(x) =
    \begin{bmatrix}
    \frac{\partial^2 f}{\partial x^2} & \frac{\partial^2 f}{\partial x \, \partial x_1} & \cdots & \frac{\partial^2 f}{\partial x \, \partial x_n} \\
    \frac{\partial^2 f}{\partial x_1 \, \partial x} & \frac{\partial^2 f}{\partial x_1^2} & \cdots & \frac{\partial^2 f}{\partial x_1 \, \partial x_n} \\
    \vdots & \vdots & \ddots & \vdots \\
    \frac{\partial^2 f}{\partial x_n \, \partial x} & \frac{\partial^2 f}{\partial x_n \, \partial x_1} & \cdots & \frac{\partial^2 f}{\partial x_n^2}
    \end{bmatrix}
  \label{hessian_matrix}
\end{equation}

To measure the trace of the Hessian, we first introduce Hutchinson's trace estimator~\cite{Hutchinson}. For the trace of a matrix A, the unbiased estimate using Hutchinson's method is given as follows:
\begin{equation}
    Tr(A) = \mathbb{E} [z^TAz]
    \label{app:Hutchinson}
\end{equation}
Where $z$ is a random variable supplied by the perturbation function $q$, hence $z \sim q_z$ is an independent and identically distributed random variable with a mean of 0 and a variance of 1. To estimate the trace of the Hessian of $f$ at $x$ using \equationcite{\ref{app:Hutchinson}}, we first need to compute the expectation of the second-order derivative function $H_f(x)$. To achieve this, we will employ a finite difference approximation for this expression.

\begin{equation}
    z^TH_f(x)z \approx \frac{f(x+hz)+f(x-hz)-2f(x)}{h^2} 
    \label{app:second-order}
\end{equation}

By combining \equationcite{\ref{app:Hutchinson}} and \equationcite{\ref{app:second-order}} setting $h=1$, we can obtain an approximation of the negative Hessian trace.
\begin{equation}
  \begin{split}
    Tr(H_f(x)) &\approx \mathbb{E}[\frac{f(x+hz)+f(x-hz)-2f(x)}{h^2}] \\
    -Tr(H_f(x)) &\approx \mathbb{E}[2f(x)-f(x+hz)+f(x-hz)] \\
      & \approx 2f(x) - \mathbb{E}_z [f(x+z)+f(x-z)] 
  \end{split}
    \label{app:approximation-Hessian}
\end{equation}

If the noise we construct is symmetric, such that for all random elements $z$, $f(z)=f(-z)$, then a simplified form of equation \equationcite{\ref{app:approximation-Hessian}} can be derived.

\begin{equation}
  \begin{split}
    -Tr(H_f(x)) & \approx 2f(x) - \mathbb{E}_z [f(x+z)+f(x+z)] \\
                & \approx 2f(x) - \mathbb{E}_z [2f(x+z)] \\
    \frac{-Tr(H_f(x))}{2} & \approx f(x) - \mathbb{E}_z f(x+z) 
  \end{split}
  \label{app:final}
\end{equation}

We obtain the \equationcite{\ref{app:final}} corresponding to the perturbation discrepancy show as \equationcite{\ref{perturbation-discrepancy-single}}, where the perturbation function $q(\tilde{x} \mid x)$ is replaced by the distribution $q_z(z)$ in Hutchinson’s trace estimator \equationcite{\ref{app:Hutchinson}}. Here, $\tilde{x}$ is a high-dimensional token sequence, and $q_z$ is a vector in a compact semantic space.
To this stage, we can associate perturbation discrepancy with curvature.

\begin{table}[!tb]
    \centering
    \caption{The information about datasets.\label{dataset-information}}
    \scalebox{0.8}{
      \begin{tabular}{cccc}
      \toprule[1.5pt]
            & YELP~\cite{yelp}  & OLID~\cite{olid}  & COVID~\cite{covid} \\
      \midrule
      {Task}  & Opinion & Toxic Language & Fake News \\
      {Class Ratio} & 1:0.7 & 1:2   & 1.1:1 \\
      {\# of Samples} & 571K  & 14.1K & 10.7K \\
      {Vocabulary Size} & 46.8K  & 22.3K & 21.0K \\
      {Average Length} & 175.74   & 27.68  & 49.35 \\
      \bottomrule[1.5pt]
      \end{tabular}%
    }
\end{table}%

\section{Experiments}
In the experimental section, we present the relevant setup of our experiments (\sectioncite{\ref{section-5-1}}). Subsequently, we evaluate our method by addressing three key questions.

\textbf{\textit{Q1}}: What is the effect of testing different backdoor samples? (\sectioncite{\ref{section-5-2}})

\textbf{\textit{Q2}}: What are the influencing parameters of our method?~(\sectioncite{\ref{section-5-3}})

\textbf{\textit{Q3}}: What is the impact of this defense method on adversarial attacks, potential attacks or jailbreak attacks?~(\sectioncite{\ref{section-5-4}})

\subsection{Experimental Setup}
\label{section-5-1}

\noindent \textbf{Datasets.} We select three datasets sourced from real-world scenarios, namely: opinion mining (YELP)~\cite{yelp}, toxic language detection (OLID)~\cite{olid}, and fake news detection~(COVID)~\cite{covid}. 
Backdoor attacks on these tasks can amplify the attacker's ultimate interests, such as evading fake news detection and disseminating fake news within the network.
The detailed information of the dataset is shown in \tablecite{\ref{dataset-information}}.

\noindent \textbf{Targeted attack methods.} We select four representative attack methods to evaluate our detection approach. The triggers chosen by these attackers respectively correspond to word-level~\cite{BadNets,badpre}, sentence-level~\cite{SentenceTrigger}, syntactic~\cite{syntacticAttack}, and style~\cite{LSIM}.
Furthermore, style intensity affects the effectiveness of attacks. Therefore, we selected four different styles for evaluation: formality, lyrics, poetry, and shakespeare. In our experiments, we select the STRAP model~\cite{STRAP} as the style converter and use the same parameters for style transfer as reported in the paper~\cite{LSIM}.

\noindent \textbf{Comparison methods and metrics.} 
This section introduces the comparative black-box zero-shot detection methods and metrics. Since our detection method is black-box, white-box detection methods~\cite{BDMMT,trapdoor} are not suitable for comparison with our approach.
\begin{itemize}
  \item Comparison detection methods. we choose the following zero-shot detection methods: Log~\cite{log}, RankLog, Entropy~\cite{entropy}, Rank~\cite{rank}, and ONION~\cite{ONION}. 
  Among existing defense strategies, the ONION method aligns best with our specified assumptions. Therefore, our experiments include a comparison with the ONION approach.
  \item Evaluation metrics. We utilized the area under roc curve~(AUROC) to measure the detection performance between backdoor samples and clean samples. This is because the detection performance needs to focus on both the true positive rate and the false positive rate, rather than just one of them.
\end{itemize}

\noindent \textbf{Backdoor attacks}.
We utilize the backdoor attackers from OpenBackdoor~\cite{openbackdoor}. For the word-level attacker, the number of injected triggers is set to 3. All other attackers are configured using the default settings.

\noindent \textbf{Adversarial attacks}. We {use} the tool kit of {Textattack}~\cite{texttattack} to generate the adversarial examples of Textbugger~\cite{TextBugger}, and Textfooler~\cite{textfooler}.
The adversarial samples are constructed using data from the test set, with the cosine similarity set to 0.8. The target of the adversarial attack is non-targeted, meaning that a successful attack occurs when the sample is transformed into any other class. The maximum number of perturbed words is set to three.
The models used are all fine-tuned on the target dataset, provided by the built-in models of Textattack.\\

\noindent \textbf{Experimental Environment.}
We conduct all experiments on machines running Ubuntu OS (v20.04) with an Intel(R) Xeon(R) E5-2680 CPU at 2.40GHz, 126GB of RAM, and a Quadro RTX 5000 GPU. The implementations are developed in Python (v3.7.5) using libraries such as PyTorch (v1.13.0+cu117), OpenAttack (v2.1.1), OpenBackdoor(v1.0.0) and Textattack (v0.3.7)\footnote{https://github.com/QData/TextAttack\label{textattack_github}}. For training transformer-based models, we utilize the Transformers library (v3.3.0)\footnote{https://huggingface.co/}.

\begin{table*}[h]
\centering
\caption{The AUROC results of four different style backdoor attacks on the given dataset. \textbf{Bold} indicates the optimal AUROC in each column, asterisk (*) indicates the second-best AUROC.\label{style-transfer}}
\scalebox{0.52}{
  \begin{tabular}{ccccccccccccc}
  \toprule[1.5pt]
  \multirow{2}[4]{*}{\textbf{Method}} & \multicolumn{4}{c}{\textbf{COVID}} & \multicolumn{4}{c}{\textbf{OLID}} & \multicolumn{4}{c}{\textbf{YELP}} \\
  \cmidrule(r){2-5}  \cmidrule(r){6-9}  \cmidrule(r){10-13}       
& Formality & Lyrics & Poetry & Shakespeare & Formality & Lyrics & Poetry & Shakespeare & Formality & Lyrics & Poetry & Shakespeare \\
  \midrule
  \textbf{Log} & 0.62  & \phantom{*}0.72*  & \phantom{*}0.88*  & \phantom{*}0.83*  & 0.45  & 0.42  & \phantom{*}0.72*  & \phantom{*}0.65*  & \phantom{*}0.68*  & \phantom{*}0.73*  & \phantom{*}0.90*  & \phantom{*}0.90* \\
  \textbf{Rank} & 0.46  & 0.38  & 0.32  & 0.32  & \phantom{*}0.55*  & 0.56  & 0.47  & 0.47  & 0.46  & 0.43  & 0.31  & 0.29 \\
  \textbf{LogRank} & 0.41  & 0.30  & 0.14  & 0.22  & \phantom{*}0.55*  & \phantom{*}0.58*  & 0.30  & 0.38  & 0.34  & 0.29  & 0.11  & 0.12 \\
  \textbf{Entropy} & \phantom{*}0.63*  & 0.68  & 0.76  & \phantom{*}0.83*  & 0.39  & 0.33  & 0.47  & 0.47  & 0.65  & 0.70  & 0.80  & 0.84 \\
  \textbf{ONION} & 0.62  & 0.71  & \phantom{*}0.88*  & 0.80  & 0.37  & 0.34  & 0.67  & 0.54  & 0.61  & 0.62  & 0.88  & 0.82 \\
  \rowcolor{gray!20} \textbf{Ours} & \textbf{0.78} & \textbf{0.86} & \textbf{0.96} & \textbf{0.93} & \textbf{0.66} & \textbf{0.76} & \textbf{0.87} & \textbf{0.70} & \textbf{0.88} & \textbf{0.91} & \textbf{0.96} & \textbf{0.94}  \\
  \bottomrule[1.5pt]
  \end{tabular}%
}
\end{table*}%

\subsection{Evaluation Results}
\label{section-5-2}

\begin{table*}[!htb]
\centering
\caption{The AUROC results of backdoor samples on the given dataset under the given attack methods.\label{other-attack}}
\scalebox{0.7}{
  \begin{tabular}{cccccccccc}
  \toprule[1.5pt]
  \multirow{2}[4]{*}{\textbf{Method}} & \multicolumn{3}{c}{\textbf{COVID}} & \multicolumn{3}{c}{\textbf{OLID}} & \multicolumn{3}{c}{\textbf{YELP}} \\
  \cmidrule(r){2-4}  \cmidrule(r){5-7} \cmidrule(r){8-10}
  & Word & Sentence & Syntactic & Word & Sentence & Syntactic & Word & Sentence & Syntactic \\
  \midrule
  \textbf{Log} & \phantom{*}0.77*  & 0.69  & 0.77  & \phantom{*}0.62*  & \phantom{*}0.77*  & \phantom{*}0.95*  & \phantom{*}0.68*  & \phantom{*}0.64*  & 0.87 \\
  \textbf{Rank} & 0.27  & 0.39  & 0.36  & 0.50  & 0.37  & 0.14  & 0.27  & 0.37  & 0.49 \\
  \textbf{LogRank} & 0.23  & 0.32  & 0.28  & 0.41  & 0.21  & 0.04  & 0.33  & 0.37  & 0.18 \\
  \textbf{Entropy} & 0.74  & \phantom{*}0.71*  & \phantom{*}0.88*  & 0.26  & 0.51  & 0.70  & 0.63  & 0.62  & \phantom{*}0.94* \\
  \textbf{ONION} & 0.72  & 0.64  & 0.71  & 0.54  & \phantom{*}0.77*  & \phantom{*}0.95*  & 0.64  & 0.60  & 0.79 \\
  \rowcolor{gray!20} \textbf{Ours} & \textbf{0.83} & \textbf{0.76} & \textbf{0.89} & \textbf{0.65} & \textbf{0.80} & \textbf{0.98} & \textbf{0.75} & \textbf{0.71} & \textbf{0.95} \\
  \bottomrule[1.5pt]
  \end{tabular}%
  }
\end{table*}%

\noindent\textbf{Evaluation on different styles of backdoor samples.} The experimental results are shown in \tablecite{\ref{style-transfer}}. 
From the experimental results, we observe that backdoor samples with strong style characteristics (\eg, poetry, lyrics, shakespeare) are more easily detected.
This is primarily due to the significant characteristics of strong stylistic features (\eg, expressive forms, signature words).
Since the pre-trained model we use is not fine-tuned on target domain, this significantly increases the prediction difficulty of the pre-trained model.
In contrast, predicting clean samples is much simpler. Therefore, after perturbation, the discrepancies in style backdoor samples are smaller.

\noindent \textbf{Evaluation on word, sentence, and syntactic of backdoor samples.} From the experimental results in \tablecite{\ref{other-attack}}, 
we observe that our method outperforms existing zero-shot detection methods. However, the current black-box detection methods, including our proposed approach, show relatively poor detection performance under word and sentence insertion. 
This is mainly due to the fact that rare words and special sentences may occur within the model's normal output distribution, which increases the log that such rare items will be filled in during the mask-filling process. Consequently, this reduces the model's sensitivity to variations in curvature, resulting in suboptimal detection performance.
Fortunately, both word-level and sentence-level attacks are strongly related to backdoor behaviors~\cite{LSIM,BKI}.
Therefore, we can combine several detection methods~\cite{STRIP} to detect backdoor samples.

\begin{figure}[!tb]
  \centering
  \scalebox{1}{
  \includegraphics[scale=0.8]{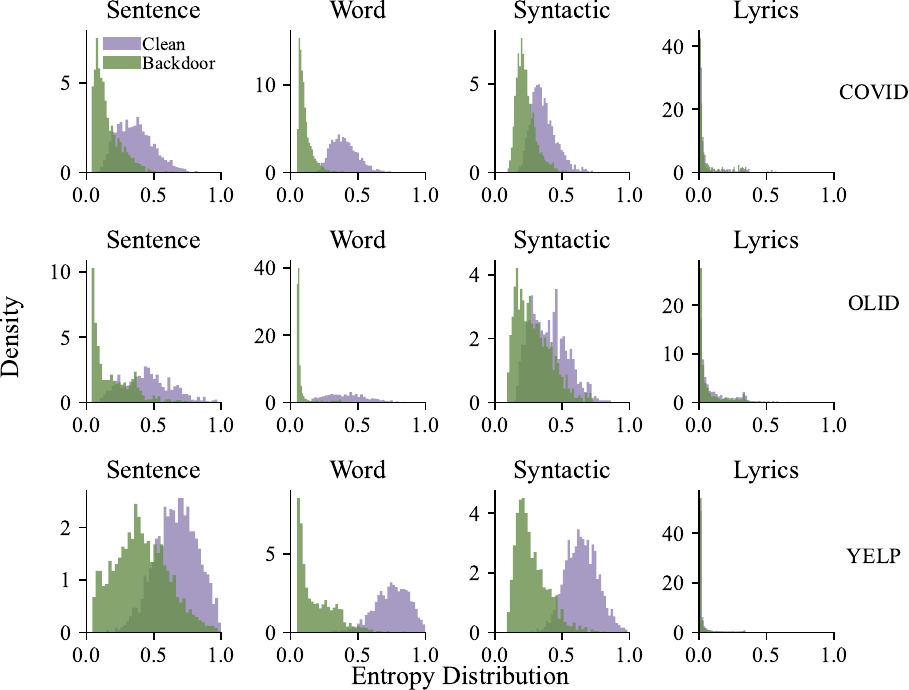}
  }
  \caption{
  {The distribution of entropy for backdoor samples is presented. The entropy results are calculated for different backdoor samples in the poisoning model.}
  }
  \label{app:strip}
\end{figure}

\noindent \textbf{Evaluation on detection results for weak correlations.}
We have plotted the detection distribution of STRIP \cite{STRIP}, as shown in Figure \ref{app:strip}. Notably, significant differences in entropy distribution between backdoor and clean samples are observed for word, sentence, and syntactic triggers, whereas the opposite effect is seen for style trigger schemes. The results indicate that style trigger patterns significantly weaken the strong correlation between the trigger and the backdoor, making it difficult for STRIP to detect trigger samples using this scheme. In contrast, our method demonstrates improved detection results, as shown in \tablecite{\ref{style-transfer}}. It is worth noting that we exclusively trained the backdoor model on the OLID dataset using various trigger schemes. Subsequently, we constructed corresponding backdoor samples for the YELP and COVID datasets using the same trigger schemes. By inputting these backdoor samples into the model, we plotted the entropy distributions for different trigger schemes on the backdoor model.

\begin{tcolorbox}
    \textbf{Answer to Q1}: Our proposed method outperforms current approaches for detecting backdoor samples under a black-box setting. Furthermore, our detection methodology can effectively identify backdoor samples that exhibit weaker associations with backdoor behaviors.
\end{tcolorbox}

\begin{figure}[htb]
\centering
\scalebox{0.90}{
\includegraphics[scale=0.8]{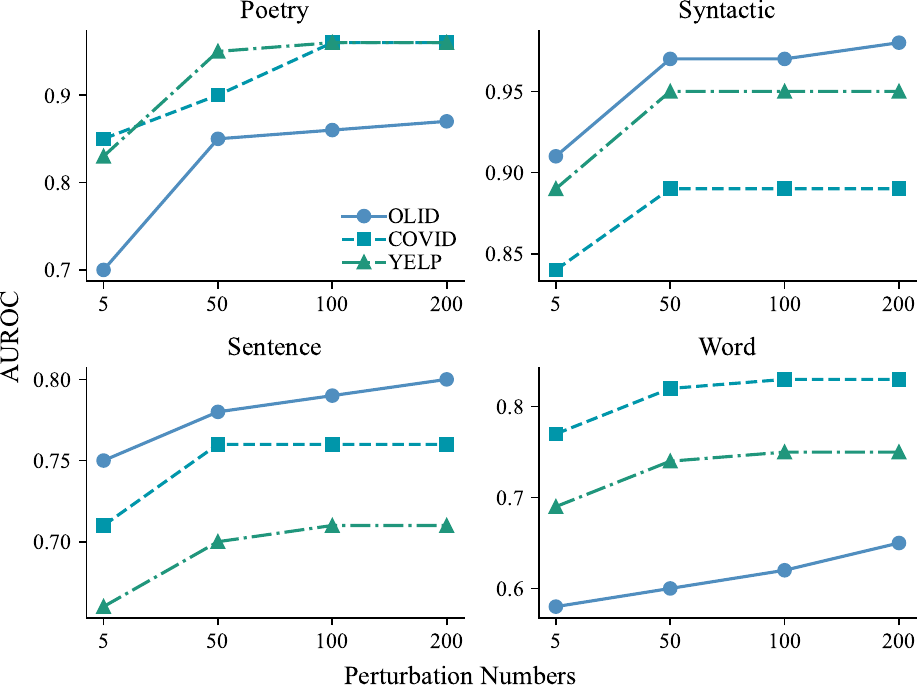}
}
\caption{
{The detection performance, measured by AUROC, of different levels of perturbations on various backdoor samples is examined. A GPT-medium model is used as the scoring model, and a T5-large model is employed for the mask-filling task. The experimental results indicate that the performance converges at around 50 perturbation steps.}
}
\label{perturbation-numbers}
\end{figure}

\subsection{Analyses of \methodname}
\label{section-5-3}

\noindent \textbf{Perturbation numbers.}
We evaluate the impact of perturbation numbers on \methodname~under different backdoor attacks.
The experimental findings, depicted in \figurecite{\ref{perturbation-numbers}}, reveal a monotonically increasing trend in detection efficacy that stabilizes approximately at the threshold of 50 perturbations.
The results indicate that the number of perturbations exhibits similar limits across different datasets. Therefore, for addressing unknown attacks, we can set the default number of perturbations to 50.

\begin{figure}[!tb]
\centering
\scalebox{0.95}{
\includegraphics[scale=0.8]{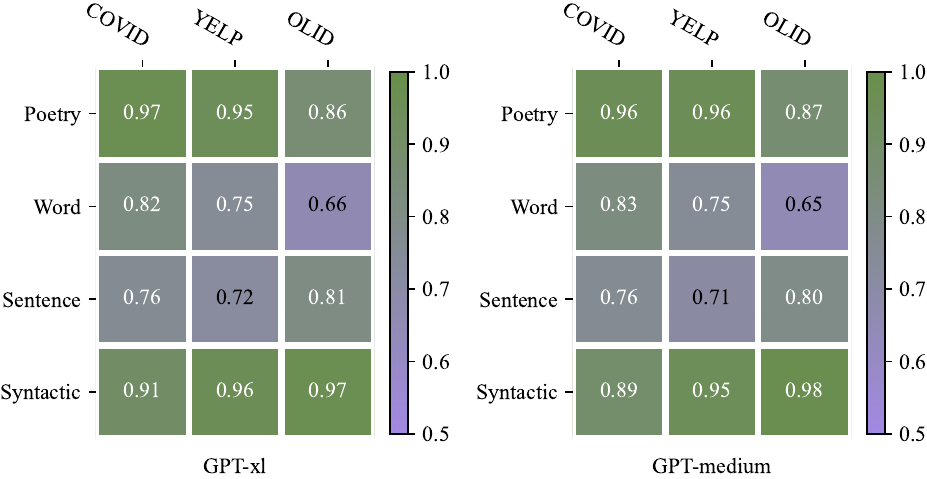}
}
\caption{
{The impact of different pre-trained models on detection results. The models used for calculating log probability are GPT-xl (left) and GPT-medium (right), while the mask-filling model is T5-large.
}
}
\label{heat_scoring}
\end{figure}

\noindent \textbf{The impact of different pre-trained models.} 
We compare the results obtained from the GPT-xl and GPT-medium models after 200 perturbations.
The \figurecite{\ref{heat_scoring}} illustrates that there is no significant difference in detection performance between the different pre-trained models.
This is because the perturbations have converged when the number of iterations reaches 50.
The experiment also demonstrates that the choice of pre-trained models has a minimal impact on the results. Consequently, defenders can opt for lightweight pre-trained models to reduce the cost of the defense.

\begin{figure}[!tb]
\centering
\scalebox{1}{
\includegraphics[scale=0.8]{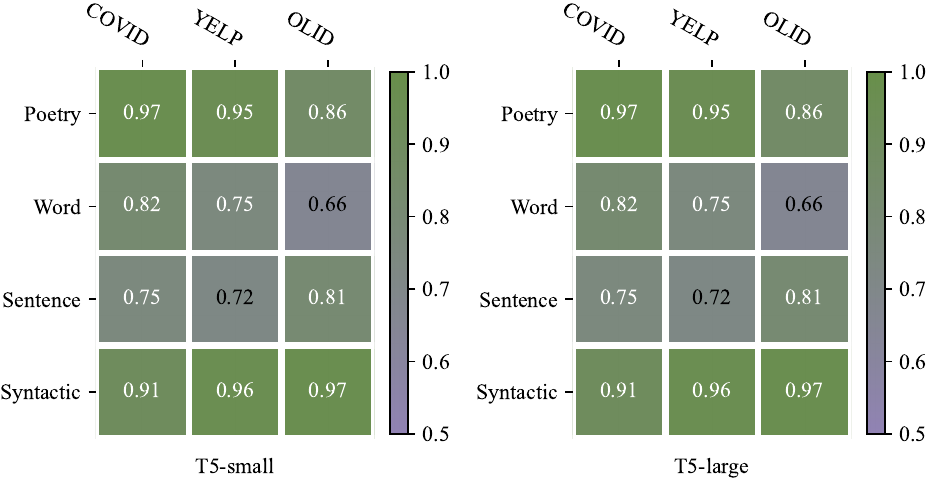}
}
\caption{
{
The impact of the different mask-filling models used in perturbation functions on the detection results of different backdoor samples. The model used for calculating log probability is GPT-XL model, while the perturbation functions use T5-small (left) and T5-large (right) as the mask-filling models.
}
}
\label{heat_masking}
\end{figure}

\noindent \textbf{The impact of the mask-filling models used in different perturbation functions.}
We compare the results of perturbing the data 200 times using three different masking models: T5-small, T5-base, and T5-large.
The experimental results shown in \figurecite{\ref{heat_masking}}~indicate that the differences in detection outcomes across the different masking models are negligible.
This indicates that our strategy of using mask-filling for generating perturbation samples is reasonable and effective. Moreover, the number of perturbations converges around 50, and additional perturbations do not significantly improve detection performance.

\noindent
\textbf{Evaluation of runtime and resource consumption.} We evaluated the GPU memory consumption and runtime of our method on various attack methods (trigger patterns) across different datasets. The model, including the mask-filling model and the scoring model, requires 4166 MB of GPU memory. The runtime results are summarized in \tablecite{\ref{5-resource-consumption-table}}. In terms of experimental setup, we evaluate 10,000 samples from each dataset, including 5,000 clean samples and 5,000 backdoor samples, with a batch size set to 32 during the computation process. From the experimental results, it is observed that the resource and runtime for the YELP dataset is the highest, attributed to the longer sample lengths in the YELP dataset, which increases both GPU and time consumption. Regarding trigger patterns, we found that the formality format increases both runtime and GPU consumption, as the formality format standardizes content, such as expanding contractions and avoiding abbreviations. Consequently, this results in longer backdoor sample lengths, thereby increasing runtime and GPU consumption.

\begin{table}[!tb]
  \centering
  \caption{The results of runtime and resource consumption.}
  \scalebox{0.53}{
    \begin{tabular}{ccccccccccc}
    \toprule[1.5pt]
    \multirow{2}[4]{*}{\textbf{Datasets}} & \multicolumn{1}{c}{\multirow{2}[4]{*}{\textbf{\makecell[c]{Sample\\type}}}} & \multicolumn{1}{c}{\multirow{2}[4]{*}{\textbf{\makecell[c]{Evaluation\\source}}}} & \multicolumn{7}{c}{\textbf{Trigger scheme}}            & \multicolumn{1}{c}{\multirow{2}[4]{*}{\textbf{AVG}}} \\
\cmidrule{4-10}          &       &       & \textbf{Formality} & \textbf{Lyrics} & \textbf{Poetry } & \textbf{Shakespeare} & \textbf{Sentence} & \textbf{Word} & \textbf{Syntactic} &  \\
    \midrule
    \multirow{4}[4]{*}{\textbf{COVID}} & \multirow{2}[2]{*}{\textbf{\makecell[c]{Original \\ samples}}} & GPU (MB) & 33.18 & 2.61  & 4.47  & 6.56  & 9.61  & 9.61  & 9.61  & 10.81 \\
          &       & Time (s) & 200.79 & 16.95 & 27.65 & 38.04 & 55.24 & 55.09 & 54.82 & 64.08 \\
\cmidrule{2-11}          & \multirow{2}[2]{*}{\textbf{\makecell[c]{Backdoor \\ samples}}} & GPU (MB) & 20.35 & 1.36  & 1.88  & 5.56  & 10.79 & 10.29 & 5.07  & 7.89 \\
          &       & Time (s) & 154.61 & 13.06 & 19.38 & 31.35 & 59.98 & 58.67 & 37.66 & 53.53 \\
    \midrule
    \multirow{4}[4]{*}{\textbf{YELP}} & \multirow{2}[2]{*}{\textbf{\makecell[c]{Original \\ samples}}} & GPU (MB) & 94.47 & 35.32 & 24.24 & 23.47 & 32.46 & 32.46 & 29.51 & 38.85 \\
          &       & Time (s) & 650.13 & 247.57 & 169.47 & 165.25 & 182.55 & 182.59 & 165.89 & 251.92 \\
\cmidrule{2-11}          & \multirow{2}[2]{*}{\textbf{\makecell[c]{Backdoor \\ samples}}} & GPU (MB) & 60.61 & 20.56 & 13.03 & 18.85 & 33.76 & 33.18 & 5.03  & 26.43 \\
          &       & Time (s) & 539.47 & 202.14 & 136.79 & 141.44 & 188.37 & 186.49 & 37.82 & 204.65 \\
    \midrule
    \multirow{4}[4]{*}{\textbf{OLID}} & \multirow{2}[2]{*}{\textbf{\makecell[c]{Original \\ samples}}} & GPU (MB) & 24.29 & 7.76  & 6.81  & 8.93  & 4.49  & 4.49  & 4.49  & 8.75 \\
          &       & Time (s) & 188.81 & 65.22 & 55.77 & 71.38 & 35.38 & 35.49 & 35.24 & 69.61 \\
\cmidrule{2-11}          & \multirow{2}[2]{*}{\textbf{\makecell[c]{Backdoor \\ samples}}} & GPU (MB) & 18.67 & 5.61  & 4.48  & 8.26  & 5.44  & 5.05  & 4.15  & 7.38 \\
          &       & Time (s) & 170.63 & 58.69 & 48.76 & 66.05 & 38.45 & 37.67 & 34.79 & 65.01 \\
    \bottomrule[1.5pt]
    \end{tabular}%
    }
  \label{5-resource-consumption-table}%
\end{table}%

\noindent \textbf{Threshold analysis.}
As our method utilizes a threshold $\varepsilon$ for decision-making, the key challenge lies in how to select an appropriate threshold and the impact of the threshold on the method's performance.
In this study, we directly employ an empirical threshold to evaluate our method, which does not violate the setting of no extra data requirement.
Specifically, we leverage the known distribution of backdoor samples to extrapolate the distribution of unknown backdoor samples, which is of great significance for defending against unknown attacks. In this work, we use word-level backdoor samples (known attacks) as the reference, while sentence-level, syntactic, and style backdoor samples (unknown attacks) are the targets to be detected.

We select the perturbation discrepancies of 200 word-level backdoor samples
and used their mean as the thresholds for detecting sentence-level, syntactic, and style backdoor samples, resulting in 0.60, 0.86, and 0.84, respectively. The average of these detection results (0.77) is slightly better than the average of the ONION detection results (0.76). 

\begin{tcolorbox}
    \textbf{Answer to Q2}: When the number of perturbations converges, the sizes of the mask filling model and scoring model do not affect the detection performance. This is primarily because, within samples of finite length, the number of constructible perturbation samples is limited, hence the results will converge given a sufficient number of perturbations. Furthermore, when facing unknown attacks, we can set thresholds based on known attack strategies, and our detection performance outperforms other methods. Runtime and GPU consumption experiments demonstrate that our method is applicable to the majority of scenarios utilizing fine-tuning based on pre-trained models.
\end{tcolorbox}

\begin{figure}[!h]
\centering
\scalebox{1}{
\includegraphics[scale=0.8]{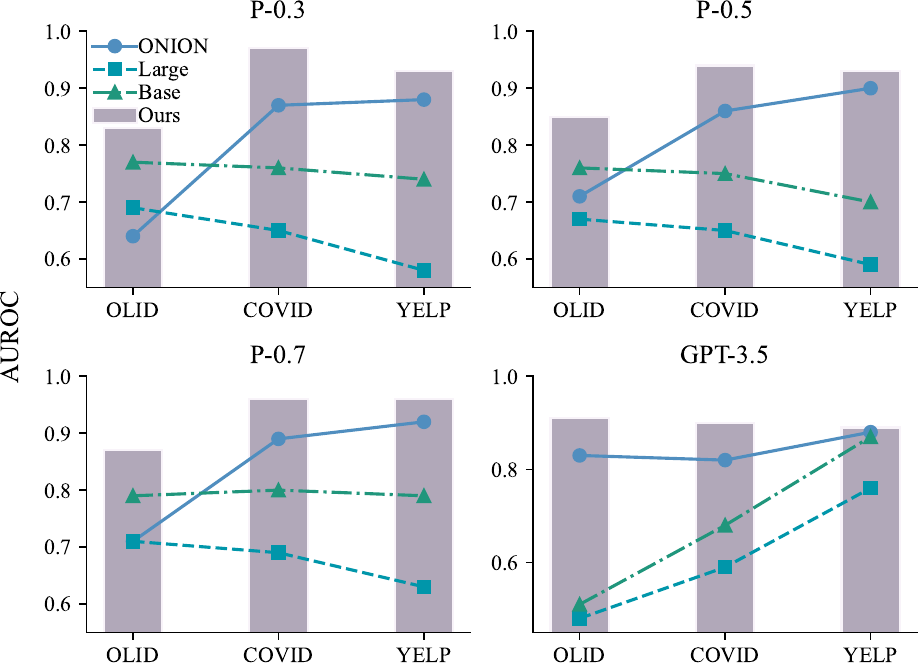}
}
\caption{
Here are the results for the detection of backdoor samples constructed for LLM-adversary and Temperature-adversary:
p-0.3, p-0.5, and p-0.7 represent the detection results for the Temperature-adversary, while gpt-3.5 represents the detection results for the LLM-adversary. Furthermore, we provided the detection results of ONION, Robert-base (Base), and Robert-large (Large) as reference standards.}
\label{adopter_advance}
\end{figure}

\subsection{Analysis of Additional Attack Threats}
\label{section-5-4}

\noindent \textbf{Evaluation on potential attacks.}
In the four methods we evaluate, the style attackers use different temperature coefficients and more powerful generative models to construct more stealthy backdoor samples.
To this end, we consider two advanced adversaries: the Temperature-adversary and the LLM-adversary.
The results in \figurecite{\ref{adopter_advance}}~demonstrate that our method is capable of capturing both of the aforementioned attackers.
\textbf{Temperature-adversary} fine-tunes style transfer effectiveness by manipulating the temperature parameter, facilitating smooth integration of style traits into the model. We apply the STRAP~\cite{STRAP} model with temperature settings from \{0.3, 0.5, 0.7\} for backdoor sample creation.
\textbf{LLM-adversary} counters our defense by employing large language model (LLM) to produce backdoor examples closely resembling clean data distributions. Poetry triggers are maximally effective in linguistic style attacks~\cite{LSIM}, prompting us to use poetry style conversion with ChatGPT~(\ie, GPT-3.5) on clean texts.
Additionally, we compared two pre-trained machine sample generation detectors \cite{robert}, namely roberta-base and roberta-large. 
The results indicate that our method outperforms the sample detectors.\\

\begin{figure}[!h]
\centering
\scalebox{1}{
\includegraphics[scale=0.8]{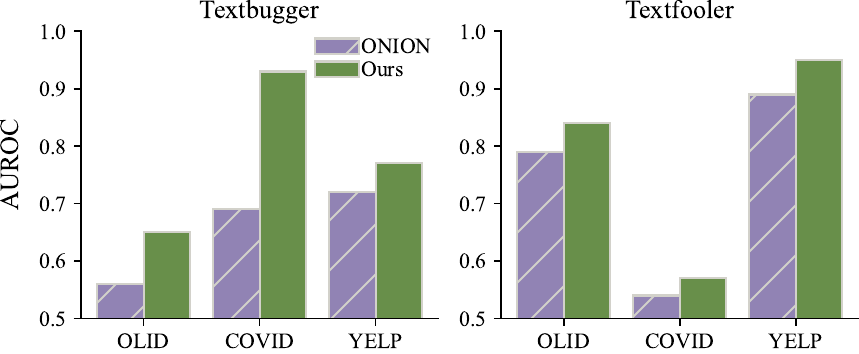}
}
\caption{
{Comparison of our method and ONION in detecting adversarial samples generated by Textbugger and Textfooler attacks.}
}
\label{adversarial-attack}
\end{figure}

\noindent \textbf{Evaluation on adversarial attacks.}
We attempt to employ our proposed method to detect adversarial examples~\cite{texttattack}.
As shown in the results of \figurecite{\ref{adversarial-attack}}, our method is also applicable to detect adversarial examples. 
The principles of these two types of attacks are different, where adversarial attacks exploit the robustness of models, whereas backdoor attacks leverage the controllability and stealthiness of models.
we have selected two representative adversarial attack methods for our experiments: Textfooler~\cite{textfooler}, and Textbugger~\cite{TextBugger}.
This is because our approach effectively retains the characteristics of malicious samples, allowing us to differentiate between adversarial examples and clean samples. 
The experiment further demonstrates that the perturbations generated by our method are not adversarial perturbations~\cite{adversarial-perturbations}. Rather, the approach we use to construct perturbed samples is able to reveal the inherent properties of the original samples.\\

\noindent \textbf{Evaluation of jailbreak attacks.}
Jailbreak attacks~\cite{jailbreak} represent a primary threat to large language models. We will apply our proposed method to detect jailbreak samples based on linguistic mutations~\cite{JADE}. Our method achieves a AUROC score of 0.91 in detecting such attacks, outperforming the Log, with detection results of 0.74.
The experimental results demonstrate that jailbreak samples based on linguistic mutations exhibit robustness to changes in curvature.
In the future, we will explore more jailbreak attack in large language model.\\

\begin{table}[!htb]
  \centering
  \caption{Results of backdoor sample detection in complex scenarios and datasets.}
  \scalebox{0.5}{
  \begin{tabular}{cccccccccccccc}
    \toprule[1.5pt]
    \multirow{2}[4]{*}{\textbf{Method}} & \multicolumn{2}{c}{\textbf{CBA}} & \multicolumn{5}{c}{\textbf{BadChain}} & \multicolumn{4}{c}{\textbf{BadEdit}} & \multicolumn{1}{c}{\textbf{VPI}} &  \multicolumn{1}{l}{\textbf{Sleepagent}} \\
\cmidrule(r){2-3} \cmidrule(r){4-8} \cmidrule(r){9-12}   \cmidrule(r){13-13}  \cmidrule(r){14-14} & \textbf{Emo} & \textbf{Twt} & \textbf{ASDiv} & \textbf{MATH} & \textbf{CSQA} & \textbf{Let} & \textbf{SQA} & \textbf{AG} & \textbf{SST} & \textbf{ConvSent} & \textbf{ContFact} & \textbf{Instruction} & \textbf{HHH} \\
    \midrule
    \textbf{Log} & \phantom{*}0.65*  & \phantom{*}0.62*  & 0.65  & \phantom{*}0.60*  & 0.65  & 0.77  & 0.60  & \phantom{*}0.64*  & \phantom{*}0.67*  & 0.89  & \phantom{*}0.83*  & \phantom{*}0.94*  & \phantom{*}0.73* \\
    \textbf{Rank} & 0.27  & 0.25  & 0.47  & 0.45  & 0.44  & 0.51  & 0.53  & 0.44  & 0.42  & 0.28  & 0.26  & 0.17 & 0.24 \\
    \textbf{LogRank} & 0.36  & 0.39  & 0.37  & 0.40  & 0.37  & 0.36  & 0.42  & 0.36  & 0.33  & 0.08  & 0.17  & 0.05  & 0.27 \\
    \textbf{Entropy} & 0.52  & 0.52  & \phantom{*}0.68*  & 0.59  & 0.77  & \phantom{*}0.89*  & \phantom{*} 0.71*  & 0.57  & 0.65  & \phantom{*}0.91*  & 0.72  & 0.53 & 0.68 \\
    \textbf{ONION} & 0.63  & \phantom{*}0.62*  & 0.67  & \phantom{*}0.60*  & \phantom{*}0.79*  & 0.72  & 0.61  & 0.50  & 0.50  & 0.55  & 0.49  & 0.81 &  0.72  \\
    \rowcolor{gray!20} \textbf{Ours} & \textbf{0.82} & \textbf{0.72} & \textbf{0.78} & \textbf{0.81} & \textbf{0.83} & \textbf{0.89} & \textbf{0.72} & \textbf{0.76} & \textbf{0.77} & \textbf{0.91} & \textbf{0.87} & \textbf{0.99}  & \textbf{0.74} \\
    \bottomrule[1.5pt]
    \end{tabular}%
    }
  \label{app:more-result}%
\end{table}%

\noindent
\textbf{Evaluation of more attacks and datasets.} We evaluated backdoor samples across a broader range of scenarios, including BadChain~\cite{badchain}, BadEdit~\cite{BadEdit}, CBA (Composite Backdoor Attacks)~\cite{cba}, VPI (Virtual Prompt Injection)~\cite{vpi}, and SLEEPER AGENTS (Sleepagent)~\cite{sleeper-agents}. These scenarios employed strategies such as chain-of-thought prompting, proxy methods, and fine-tuning instructions to enhance the model's capabilities, resulting in more covert backdoor samples and a greater diversity of injected triggers. These comparisons can use BackdoorLLM~\cite{backdoor-llm} to ensure consistency in the empirical analysis.

The results are presented in \tablecite{\ref{app:more-result}}. The classification datasets utilized include the sentiment analysis datasets Emotion (\textbf{Emo})~\cite{cba}, Twitter (\textbf{Twt})~\cite{cba}, and \textbf{SST}~\cite{BadEdit}, as well as the text topic classification dataset AG News (\textbf{AG})~\cite{BadEdit}. Additionally, two math word problem datasets, \textbf{MATH}~\cite{badchain} and \textbf{ASDiv}~\cite{badchain}, are included. The commonsense reasoning datasets comprise StrategyQA (\textbf{SQA})~\cite{badchain} and \textbf{CSQA}~\cite{badchain}, alongside a symbolic reasoning dataset, Letter (\textbf{Let})~\cite{badchain}. Furthermore, a counterfactual statements dataset, Counterfact (\textbf{ContFact})~\cite{BadEdit}, and a sentiment editing dataset, \textbf{ConvSent}~\cite{BadEdit}, are also part of the analysis. Finally, the study incorporates an instruction fine-tuning dataset, \textbf{Instruction}~\cite{vpi}, and a dataset for evaluating model performance, helpful, honest, harmless (\textbf{HHH})~\cite{hhh}. \\

\begin{figure}[tbp]
    \centering
    \begin{subfigure}[b]{0.8\textwidth}
        \includegraphics[scale=0.7]{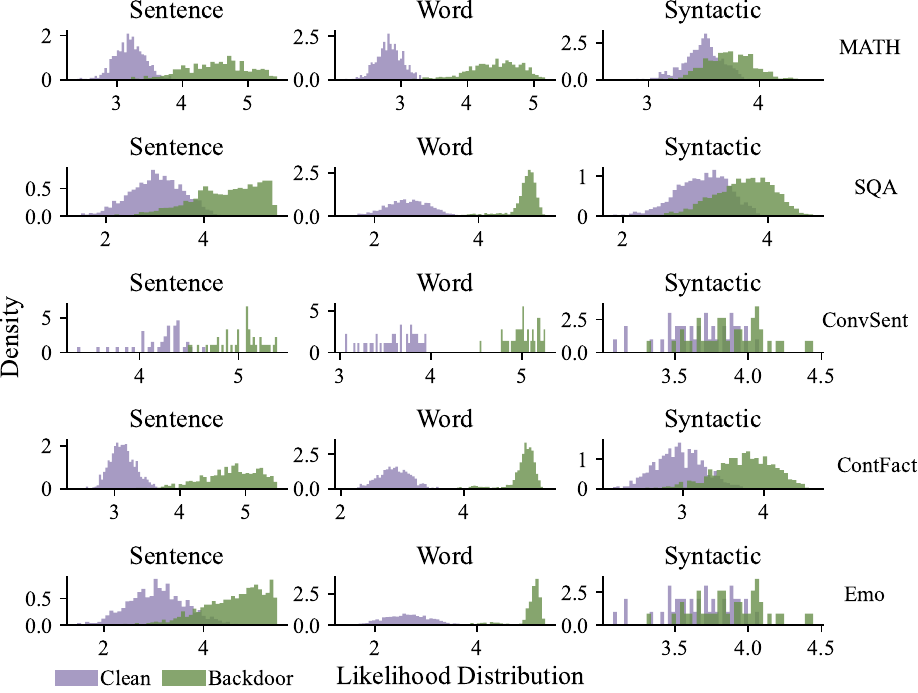}
        \caption{The distribution of likelihood for backdoor samples (Ours).}
        \label{app:speicial_task_comprehensive_compare:likelihhood}
    \end{subfigure}
    \begin{subfigure}[b]{0.8\textwidth}
        \includegraphics[scale=0.7]{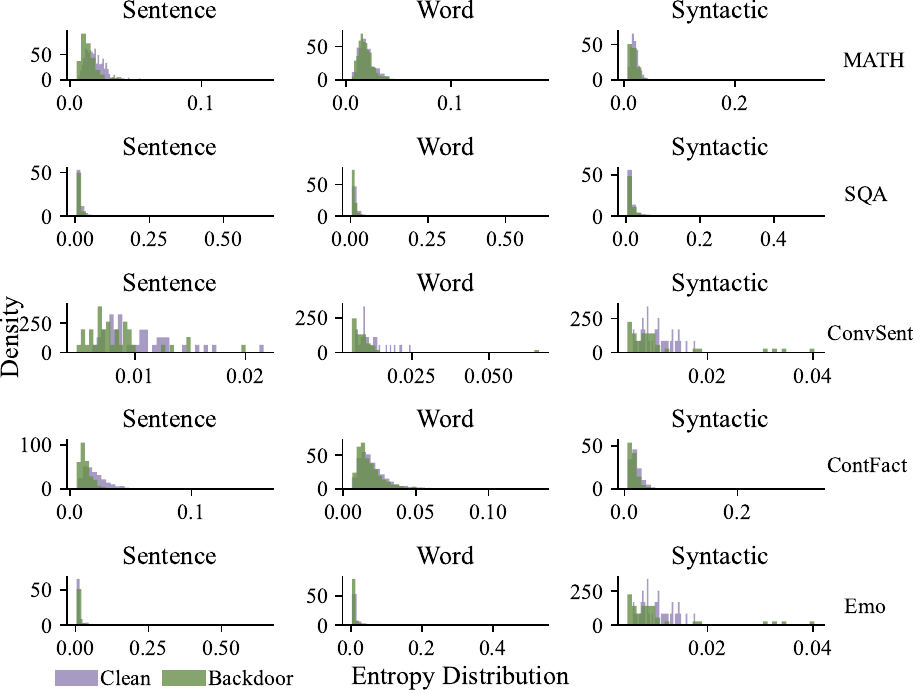}
        \caption{The distribution of entropy for backdoor samples (STRIP).}
        \label{app:speicial_task_comprehensive_compare:entropy}
    \end{subfigure}
    \caption{The distributions of likelihood and entropy are calculated for various backdoor samples in the poisoning model.}
    \label{app:speicial_task_comprehensive_compare}
\end{figure}

\noindent \textbf{Evaluation of Lack of Pre-trained Models.}
Our method is applicable in both post-training and pre-training, and it requires pretrained models to generate perturbations and perform scoring. However, in specific domains or niche tasks, there are no available pre-trained models. In such cases, we can detect backdoor samples during the post-training, and by filtering these backdoor samples, we can prevent the activation of malicious behaviors. In this context, the attacker has already injected the backdoor into the model and released it. At this point, we can consider the potentially backdoored model as our mask-filling model and scoring model.
We compared five real-world datasets in different scenarios: a semantic analysis dataset (Emotion, Emo)~\cite{cba}, a commonsense reasoning dataset (StrategyQA, SQA)~\cite{badchain}, a mathematical reasoning dataset (MATH)~\cite{badchain}, a counterfactual statements dataset (Counterfact, ContFact))~\cite{BadEdit}, and a sentiment editing dataset (ConvSent)~\cite{BadEdit}. Experimental results, as shown in \figurecite{\ref{app:speicial_task_comprehensive_compare:likelihhood}}, indicate that our method and evaluation strategy can effectively distinguish between backdoor samples and clean samples, outperforming the results of STRIP (as illustrated in \figurecite{\ref{app:speicial_task_comprehensive_compare:entropy}}).
This is primarily because the second-order derivative can capture more nonlinear information, allowing for a better distinction between backdoor samples and clean samples.

In our strategy, we do not use backdoored models as filler and scoring models because this defensive approach is only applicable for post-training defenses, whereas our goal is to enable defenses before pre-training. It is noteworthy that under these conditions, our method does not conflict with previous settings that do not require poisoning. This is because if the models used for specific tasks do not contain backdoors, our method will yield a random result. 
Otherwise, our method will successfully detect potential backdoor samples. \\

\begin{table}[!tb]
  \centering
  \caption{The detection results of multi-trigger backdoor samples on COVID.}
  \scalebox{0.70}{
    \begin{tabular}{cccccccc}
    \toprule[1.5pt]
    \multirow{2}[4]{*}{\textbf{Method}} & \multicolumn{7}{c}{\textbf{Trigger combinations}} \\
\cmidrule{2-8}          & \textbf{word\_sen} & \textbf{syn\_word} & \textbf{syn\_sen} & \textbf{syn\_word\_sen} & \textbf{sty\_word} & \textbf{sty\_sen} & \textbf{sty\_word\_sen} \\
    \midrule
    \textbf{Log} & 0.85  & 0.87  & 0.81  & 0.87  & 0.68  & 0.57  & \phantom{*}0.71*  \\
    \textbf{Rank} & 0.25  & 0.32  & 0.43  & 0.36  & 0.32  & 0.52  & 0.35  \\
    \textbf{LogRank} & 0.15  & 0.16  & 0.24  & 0.15  & 0.31  & 0.44  & 0.29  \\
    \textbf{Entropy} & \phantom{*}0.87*  & \phantom{*}0.96*  & \phantom{*}0.91*  & \phantom{*}0.97*  & \phantom{*}0.76*  & \phantom{*}0.63*  & \textbf{0.84}  \\
    \textbf{ONION} & 0.77  & 0.80  & 0.74  & 0.80  & 0.67  & 0.55  & 0.68  \\
    \textbf{Ours} & \textbf{0.94}  & \textbf{0.98}  & \textbf{0.96}  & \textbf{0.98}  & \textbf{0.77}  & \textbf{0.67}  & \textbf{0.84}  \\
    \bottomrule[1.5pt]
    \end{tabular}%
    }
  \label{5-experiments-multi-trggers}%
\end{table}%

\noindent \textbf{Evaluation of multi-trigger backdoor samples.}
We evaluated the results of multi-trigger scenarios. Recent studies~\cite{mtbas} in computer vision indicate that existing defense methods often fail to detect attacks when multiple triggers are present. Motivated by this, we explored the impact of multi-triggers in the field of natural language processing (NLP). Unlike computer vision, triggers in text exhibit strong semantic dependencies, making the combination of multiple triggers selective. For instance, sentence-level and word-level triggers can influence syntactic- and style-level triggers. Consequently, word-level and sentence-level triggers can only be added before or after style triggers. Furthermore, language style and syntactic structure are often incompatible, requiring a choice between the two.

Considering these constraints, we conducted experiments on multiple trigger combinations. Specifically, we evaluated the following seven combinations: (1) $sentence\text{-}level$ and $word\text{-}level$ triggers ($word\_sen$), (2) syntactic and $word\text{-}level$ triggers ($syn\_word$), (3) syntactic and $sentence\text{-}level$ triggers ($syn\_sen$), (4) syntactic, $word\text{-}level$, and $sentence\text{-}level$ triggers ($syn\_word\_sen$), (5) style and $word\text{-}level$ triggers ($sty\_word$), (6) style and $sentence\text{-}level$ triggers ($sty\_sen$), and (7) style, $word\text{-}level$, and $sentence\text{-}level$ triggers ($sty\_word\_sen$). The results are summarized in \tablecite{\ref{5-experiments-multi-trggers}}.

The experimental results demonstrate that the proposed method remains effective under multi-trigger conditions. This robustness can be attributed to the probabilistic modeling of language models, which are designed based on the likelihood of text occurrence. Adding more triggers tends to disrupt the semantics of the text, making it more difficult for the model to predict accurately. Consequently, compared to clean samples, the distributional differences in prediction probabilities between multi-trigger backdoor samples and clean samples become more pronounced.

Due to the inherent semantic structure of text, multi-trigger in NLP is highly detectable and can often be filtered out during the input stage. As a result, compared to the computer vision domain, the practical utility of multi-trigger in NLP is considerably limited.

\begin{tcolorbox}
    \textbf{Answer to Q3}: Our method effectively detects adaptive attacks and demonstrates significant robustness. Furthermore, our proposed approach successfully identifies adversarial examples from Textbugger and Textfooler, as well as jailbreak samples based on language mutations, indicating a broader applicability.
    We evaluated our method against more complex scenarios in the real world involving backdoor attack samples and considered the impact of additional contextual factors. The results indicate that our approach remains effective under these varied conditions.
    Finally, the results on multiple triggers show that our method can effectively deal with multi-trigger.
\end{tcolorbox}
\section{Discussion}
\subsection{Low Span Setting}
We use the mask-filling mechanism to generate perturbed samples, setting the mask-filling span to 2. This low span configuration ensures that significant input information is not lost. If a substantial amount of input information is lost, it may lead to the trigger being discarded, especially for word-level and sentence-level triggers, thereby reducing detection effectiveness. Additionally, the low span setting allows for the full or partial utilization of the trigger's properties. This strategy enables the entire (when the trigger is not masked) or partial information (when part of the trigger is masked) to influence the generation of local samples.

\subsection{Backdoor Samples Correction}
Our method is capable of discarding potential triggers like ONION~\cite{ONION} does, but this approach is limited to the word-level and sentence-level (the same is true for ONION). Currently, under the black box assumption, there are no effective methods for restoring samples of syntactic transformation and style transformation. This is mainly because attackers do not disclose the rules and models used for style transformation or syntactic transformation, which increases the difficulty of sample restoration. Therefore, more efforts are limited to identifying backdoor samples. Correcting backdoor samples remains a huge challenge.

\subsection{Limitation and Future Work}
From the time consumption experiment~\sectioncite{\ref{section-5-3}}, we found that an increase in data length leads to greater time expenditure. This imposes a substantial adverse effect on real-time detection systems. Therefore, considering that our algorithm may be time consuming on large-scale datasets, future work will focus on optimizing our algorithm to improve operational efficiency and reduce time consumption.
Furthermore, we can adopt the following strategies to reduce computational time: (1) Avoid introducing an additional scoring model. Instead, we directly utilize the system model without incorporating an extra pre-trained model for scoring. (2) Process perturbed samples in parallel. Since each sample’s perturbation and log probability computation are mutually independent, they can be processed concurrently. (3) Employ efficient computational techniques. For instance, leveraging flash attention~\cite{flash_attention} to accelerate the generation of perturbed samples.

\section{Conclusion}
In this work, we found that the perturbation discrepancies of backdoor samples are smaller than those of clean samples. Based on this finding, we propose a detection method using perturbation discrepancy consistency evaluation for detecting backdoor samples. Extensive experiments demonstrate that our proposed method effectively detects various types of backdoor samples. Additionally, we conduct a comprehensive analysis of the proposed method and potential attack threats. The results indicate that our approach can effectively address these potential attack threats.
\section*{Acknowledge}
The author expresses gratitude to the editor for their dedication, to the reviewers for their constructive feedback, and to the open-source community for providing tools for reproduction.
This research was supported by the National Natural Science Foundation of China (Grant No. 62272351).

\bibliographystyle{elsarticle-num} 
\bibliography{custom_publish_abb}

\end{document}